\documentstyle[twocolumn,seceq]{jpsj}
\def\runtitle{Critical Dimerization Strength of the Quarter-Filled 
$t$$-$$J$ Model}
\def\runauthor{Satoshi {\sc Nishimoto} and Yukinori {\sc Ohta}}
\title{Critical Dimerization Strength of the Quarter-Filled 
$t$$-$$J$ Model
}
\author{Satoshi {\sc Nishimoto}$^{1}$ and Yukinori {\sc Ohta}$^{1,2}$}
\inst
{$^1$ Graduate School of Science and Technology, Chiba University, 
Inage-ku, Chiba 263-8522\\
$^2$Department of Physics, Chiba University, Inage-ku, Chiba 263-8522}
\recdate{May 1, 1998}
\abst
{An exact-diagonalization technique on small clusters is used to study 
the ground state of the dimerized $t$$-$$J$ model at quarter filling.  
The equal-time charge and spin correlations, charge and spin gaps, 
and binding energy are calculated for the two-dimensional lattice 
with a spatial dimerization pattern corresponding to organic conductors 
$\kappa$-(BEDT-TTF)$_2$X.  We show that competition between 
the effects of two types of dimerization, i.e., dimerization in 
the hopping integral and dimerization in the exchange interaction, 
leads to the insulator-superconductor transition in the ground state 
of the model.  The phase diagram is thereby presented.}
\kword
{$t$$-$$J$ model, dimerization, spin gap, charge gap, 
insulator-superconductor transition, BEDT-TTF}
\begin{document}
\sloppy
\maketitle

It has been emphasized\cite{rf:1} that quasi-two-dimensional (2D) 
organic conductors $\kappa$-(BEDT-TTF)$_2$X have some characteristics 
similar to cuprate superconductors;  apart from 2D nature of 
electron conduction, strong electron correlations, and Mott-insulator 
to superconductor transition, recent $^{13}$C-NMR\cite{rf:2,rf:3,rf:4} 
and specific-heat measurements\cite{rf:5,rf:6,rf:7} have suggested 
that the superconducting phase is anisotropic with $d$-wave--like nodes 
in the gap function and the behavior of $(T_1T)^{-1}$ is somewhat 
similar to the spin-gap phenomena observed in cuprate 
superconductors.\cite{rf:8,rf:9}  Also noticeable is the very short 
coherence length of $\xi_0$$=$$25$$\sim$$30$\AA,\cite{rf:10} which is 
comparable to that for cuprate superconductors.  

Motivated by these intriguing experimental data on the materials and 
some theoretical calculations on the dimerized Hubbard models,\cite{rf:11} 
we here study the effects of dimerization on the ground state of the 
quarter-filled $t$$-$$J$ model, another representative model for strong 
electron correlations, by focusing on its insulator-superconductor 
transition.  We will show that due to competition between two types 
of dimerization the ground-state phase of the model is divided into 
a Mott-insulating phase and a singlet-pairing phase; i.e., there is 
a critical dimerization strength.  Although applicability of the model 
to the organic compounds is not yet fully worked out, we believe that 
the results shown here will help one understand a possible mechanism 
of superconductivity in strongly correlated electron systems such as 
$\kappa$-(BEDT-TTF)$_2$X.  

The dimerized $t$$-$$J$ model we study is defined by the Hamiltonian 
\begin{eqnarray}
&H&=-t_1\sum_{\langle ij \rangle\sigma}
(\hat{c}^\dagger_{i\sigma}\hat{c}_{j\sigma}+{\rm H.c.})
-t_2\sum_{\langle kl\rangle\sigma}(\hat{c}^\dagger_{k\sigma}
\hat{c}_{l\sigma}+{\rm H.c.}) \nonumber \\
&+&J_1\sum_{\langle ij\rangle}({\bf S}_i\cdot{\bf S}_j-\frac{n_i n_j}{4})
+J_2\sum_{\langle kl\rangle}({\bf S}_k\cdot{\bf S}_l-\frac{n_k n_l}{4})\
\end{eqnarray}
where $\hat{c}^\dagger_{i\sigma}$$=$$c^\dagger_{i\sigma}(1$$-$$n_{i-\sigma})$ 
is the constrained electron-creation operator at site $i$ and spin $\sigma$ 
$(=\uparrow,\downarrow)$, ${\bf S}_i$ is the spin-$\frac{1}{2}$ operator, 
and $n_i$ is the electron-number operator; we refer to the fermionic 
particle as `electron', which corresponds to the hole in the real organic 
compounds.  $\langle ij\rangle$ stands for nearest-neighbor bonds with 
parameters $t_1$ and $J_1$ and $\langle kl\rangle$ for those with 
parameters $t_2$ $(\ge$$t_1)$ and $J_2$ $(\ge$$J_1)$.  
We assume the lattice of a dimerization pattern simulating the 2D 
conducting plane of $\kappa$-(BEDT-TTF)$_2$X compounds (see Fig.~1(a)).  
Note that the model tends to the usual square-lattice $t$$-$$J$ model 
when there is no dimerization ($t_1$$=$$t_2$ and $J_1$$=$$J_2$), of which 
much study has been made,\cite{rf:12,add:1} whereas in the limit of strong 
dimerization, the model represents an assembly of isolated dimers.  
We retain the relations between parameters $t$ and $J$ obtained from 
perturbation, i.e., $J_1$$=$$4t^2_1/U$ and $J_2$$=$$4t^2_2/U$, in order 
to reduce the number of parameters, where $U$ is the corresponding on-site 
Hubbard interaction.  We thereby keep a relation $J_1/J_2$$=$$(t_1/t_2)^2$.  
We thus have three independent parameters, and if we take $t_1$ as a unit 
of energy, then we have two, for which we will take parameters representing 
$t$-dimerization and $J$-dimerization (of which a specific definition is 
given below).  Here we restrict ourselves to the case of quarter filling 
to simulate the situation of the organic compounds.  

We introduce two types of dimerization: one is the dimerization of 
hopping integral, which we call $t$-dimerization, and the other is the 
dimerization of exchange interaction, which we call $J$-dimerization.  
We define a parameter 
\begin{equation}
\tilde{t}_d=\frac{t_2-t_1}{t_1}
\end{equation}
for the strength of $t$-dimerization and a parameter 
\begin{equation}
\tilde{J}_d=\frac{J_2-J_1}{t_1}
\end{equation}
for the strength of $J$-dimerization.  We also take $J_2/t_2$ as a measure 
of the strength of $J$-dimerization because if we keep $\tilde{t}_d$ constant 
$(>0)$ then $J_2/t_2\propto \tilde{J}_d$.  We note that the $t$-dimerization 
has the effect leading to a repulsive interaction among electrons that acts 
when different spins ($\uparrow$ and $\downarrow$) come in a single dimer 
and that the $J$-dimerization has the effect promoting the spin-singlet 
formation between spins coming in a single dimer.  
Thus, the competition between the effects of these two dimerization may 
result in the following situation as illustrated in Fig.~1(b); when 
$t$-dimerization is dominant, the system can be a Mott insulator 
(because we have an effective half-filled band with the `on-site' 
repulsion $U_{\rm dimer}$\cite{rf:11}), when $J$-dimerization is dominant, 
the system can be a singlet-pairing state, and when $J_2/t_2$ is very large, 
the system will be phase separated.  
\begin{figure}
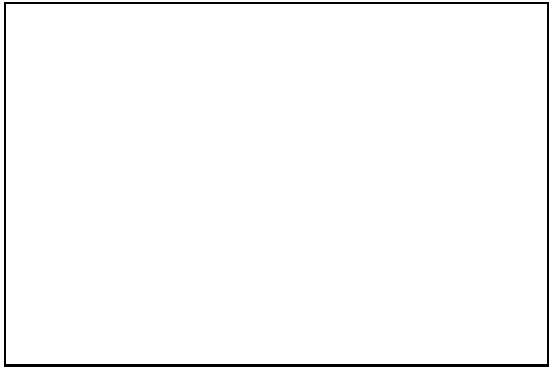

\figureheight{4.6cm}
\caption{(a) A 2D lattice structure of $\kappa$-(BEDT-TTF)$_2$X.  
The nearest-neighbor bonds have either $t_1$ and $J_1$ (thin solid line) 
or $t_2$ and $J_2$ (bold line).  The unit cells, each of which contains 
four sites, are indicated by dashed lines.  
(b) Schematic 1D representation of the electronic states: from top to 
bottom, Mott insulator, singlet-pairing state, and phase separation 
are illustrated.}
\label{fig:1}
\end{figure}

In the following, we will present numerical evidences that the competition 
indeed leads to the insulator-superconductor transition in the ground-state 
phase diagram of the dimerized $t$$-$$J$ model.  We employ a numerical 
exact-diagonalization technique on small clusters.  We use clusters of the 
size $1$$\times$$4$ and $2$$\times$$2$ unit cells (16 sites) with periodic 
boundary conditions (see Fig.~1 (a)).  We find that nearly the same results 
are obtained for these two clusters, so that hereafter we will show the 
results only for the cluster of $2$$\times$$2$ unit cells.  
\begin{figure}
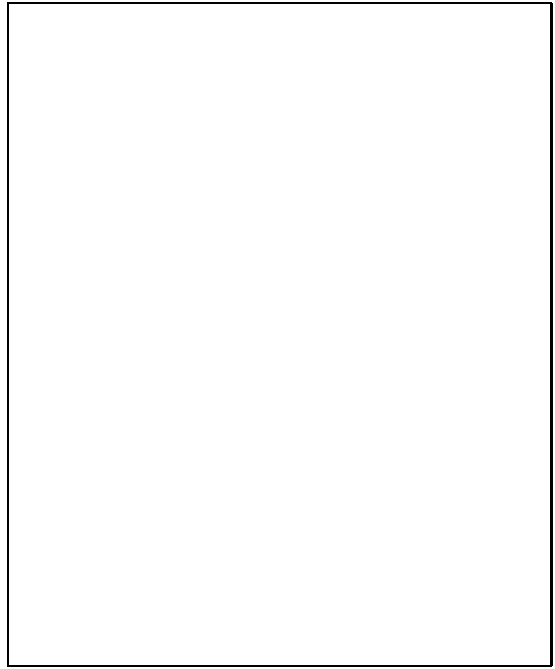

\figureheight{8.6cm}
\caption{Charge correlation $\langle n_in_j\rangle$ (left panel) and 
spin correlation $\langle S_i^zS_j^z\rangle$ (right panel) for the 
dimerized $t$$-$$J$ model.  Intra-dimer correlations are indicated by 
full squares.  Topmost two panels are for the cases without 
dimerization.}
\label{fig:2}
\end{figure}

Let us first discuss the equal-time charge and spin correlations.  
In Fig.~2, we show the calculated results for the charge 
correlation $\langle n_in_j\rangle$ and spin correlation 
$\langle S^z_iS^z_j\rangle$, where $\langle\cdots\rangle$ denotes the 
ground-state expectation value.  For comparison, we also show the results 
for the $t$$-$$J$ cluster without dimerization 
($\tilde{t}_d$$=$$\tilde{J}_d$$=$$0$) in the topmost panels of Fig.~2.  
It is first of all evident in the figures that the effect of 
dimerizations indeed bring strong impact on the intra-dimer charge and 
spin correlations.  
We find that, in the small $J_2/t_2$ region where $J$-dimerization 
is weak, the intra-dimer charge correlation becomes closer to zero, 
indicating a tendency that only one electron presents in a dimer.  
This suggests that the effective repulsive interaction presents 
in a dimer.  With increasing $t$-dimerization ($\tilde{t}_d$) this 
tendency appears to be much enhanced as clearly seen in Fig.~2.  
On the other hand, in the large $J_2/t_2$ region where the $J$-dimerization 
is large, the intra-dimer charge correlation is approaching the value 1/4 
and the intra-dimer spin correlation is decreasing to $-$1/16 as clearly seen 
in Fig.~2.  This result indicates that two electrons with opposite spin 
come in a single dimer when $J$-dimerization is large.  
Then, in the region of intermediate strength of $J_2/t_2$, there appears 
a state where the charge correlations among different sites have nearly 
the same value $\sim$$0.5/4$ and thus there are rather small spatial 
correlations.  
We note however that even in this region the nearest-neighbor spin 
correlations are still significantly antiferromagnetic.  
\begin{figure}
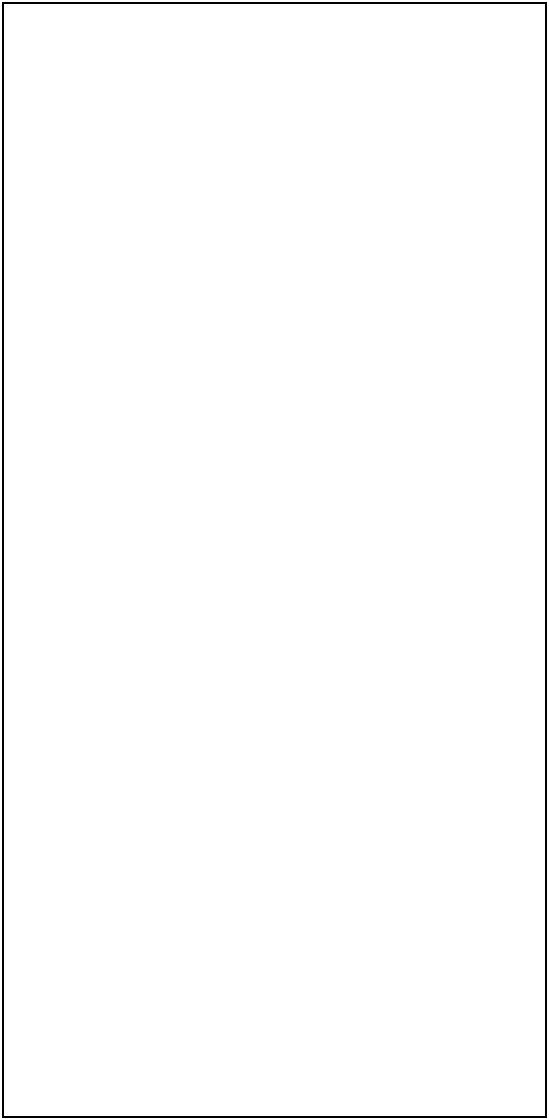

\figureheight{14.6cm}
\caption{(a) Single-particle gap $\Delta_{c1}/t_2$, 
(b) charge gap $\Delta_{c2}/t_2$, 
(c) spin gap $\Delta_{s}/t_2$, and 
(d) binding energy $\Delta_{B}/t_2$ as a function of $J_2/t_2$.}
\label{fig:3}
\end{figure}

Next let us examine the single-particle gap and charge gap.  For the 
16-site cluster at quarter filling the single-particle gap may be 
defined by 
\begin{eqnarray}
\Delta_{c1}=\frac{1}{2}\Bigl[ [E_{\rm GS}(5,4)&-&E_{\rm GS}(4,4)] \nonumber \\
&-&[ E_{\rm GS}(4,4)-E_{\rm GS}(3,4)] \Bigr]  \
\end{eqnarray}
where $E_{\rm GS}(N_\uparrow,N_\downarrow)$ is the ground-state energy for 
the cluster of $N_\uparrow$ up-spin and $N_\downarrow$ down-spin electrons.  
The calculated results are shown in Fig.~3 (a).  
When there is no dimerization, one might expect that the system is metallic, 
i.e., $\Delta_{c1}$$=$$0$ in the entire region of $J_2/t_2$ (except in the 
region of phase separation).  We see however that the obtained value for 
$\Delta_{c1}$ is finite and it increases with increasing of $J_2/t_2$.  
The reason is, apart from an obvious finite-size effect of small clusters, 
that $\Delta_{c1}$ reflects the effect of electron pairing, even in the 
charge-gapless region (as discussed below).  
When there is dimerization, $\Delta_{c1}$ is finite even at $J_2/t_2$$=$$0$ 
and rapidly increases with increasing the strength of $t$-dimerization.  
We see that at constant $\tilde{t}_d$ and with increasing $J$-dimerization 
$J_2/t_2$, the gap first decreases, reaches a minimum, and then increases, 
i.e., there appears a minimum value in $\Delta_{c1}$.  The value of 
$J_2/t_2$ at which this minimum occurs is defined as $(J_2/t_2)_c$.  
The minimum position $(J_2/t_2)_c$ increases with increasing $t$-dimerization.  

The charge gap defined by 
\begin{eqnarray}
\Delta_{c2}=\frac{1}{2}\Bigl[ [E_{\rm GS}(5,5)&-&E_{\rm GS}(4,4)] \nonumber \\
&-&[E_{\rm GS}(4,4)-E_{\rm GS}(3,3)] \Bigr]  \
\end{eqnarray}
may also be introduced.  The calculated results are shown in Fig.~3 (b).  
When there is no dimerization, we find the result $\Delta_{c2}$$<$$0$, 
contrary to a simple expectation.  
This may however be understood if we assume that the attractive interaction 
between pairs (or some tendency toward phase separation), of which effect 
$\Delta_{c2}$ reflects, exists in the singlet-pairing region.  
When there is dimerization, the charge gap $\Delta_{c2}$ opens for 
$\tilde{t}_d$$>$$0$ and increases rapidly with increasing $\tilde{t}_d$.  
With increasing $J_2/t_2$, the gap decreases first and at some $J_2/t_2$ 
value the rate of the decrease changes, and at the same time $\Delta_{c2}$ 
changes sign where the gap closes.  We may write this $J_2/t_2$ value as 
$(J_2/t_2)_c$ because the value agrees with the value of $(J_2/t_2)_c$ 
defined by using $\Delta_{c1}$.  Above $(J_2/t_2)_c$, the gap decreases as 
$J_2/t_2$ increases, which may again be due to the effect of attractive 
interaction between pairs.  

Let us now consider the spin gap, which is defined by 
\begin{equation}
\Delta_{s} = E_{\rm GS}(5,3)-E_{\rm GS}(4,4).  
\end{equation}
The calculated results are shown in Fig.~3 (c).  
We find that the spin gap opens at $J_2/t_2$$\sim$$(J_2/t_2)_c$ and 
increases rapidly with increasing $J_2/t_2$ (if we assume that a small 
$\Delta_s$$>$$0$ at $J_2/t_2$$<$$(J_2/t_2)_c$ is a finite-size effect).  
We also find that with increasing $\tilde{t}_d$ the size of the gap 
decreases and the value $(J_2/t_2)_c$ increases.  
Here we should emphasize that the $(J_2/t_2)_c$ values estimated from 
the behaviors of $\Delta_{c1}$, $\Delta_{c2}$, and $\Delta_s$ are all 
consistent and have the same $\tilde{t}_d$ dependence.  
Thus it seems quite reasonable to assume that the spin gap $\Delta_{s}$ 
is finite in the entire charge-gapless region, where the electrons 
form a bound state as shown below.  

The binding energy may be estimated by 
\begin{eqnarray}
\Delta_{\rm B}=[E_{\rm GS}(5,5)&-&E_{\rm GS}(4,4)]   \nonumber \\
&-&2[E_{\rm GS}(5,4)-E_{\rm GS}(4,4)] \
\end{eqnarray}
where $\Delta_{B}$ is negative if the state of two electrons minimize 
its energy by forming a bound state.  A possibility of superconductivity 
may then be indicated.  The calculated result for $\Delta_{\rm B}$ is 
shown in Fig.~3 (d).  We find that $|\Delta_{\rm B}|$ is small for 
$J_2/t_2$$<$$(J_2/t_2)_c$ but starts to increase at around $(J_2/t_2)_c$, 
which is consistent with the expectation that the bound state of 
electrons is formed in the region of charge-gapless and spin-gapful phase.  
We also note that, with increasing $t$-dimerization, the binding energy 
is suppressed and the value of $(J_2/t_2)_c$ increases accordingly.  
We should however be careful because when the value of $J_2/t_2$ is 
too large, the system will be phase separated.  We estimate the parameter 
region of phase separation by examining if the value of the 
compressibility becomes negative.\cite{rf:13}  It is known that the phase 
separation actually occurs at around $J_2/t_2$$\simeq$$2.5$$-$$3$ when 
there is no dimerization.\cite{rf:14}  We thus examine its dependence 
on the $t$- and $J$-dimerizations; the obtained results are summarized 
in the phase diagram shown below.  
\begin{figure}
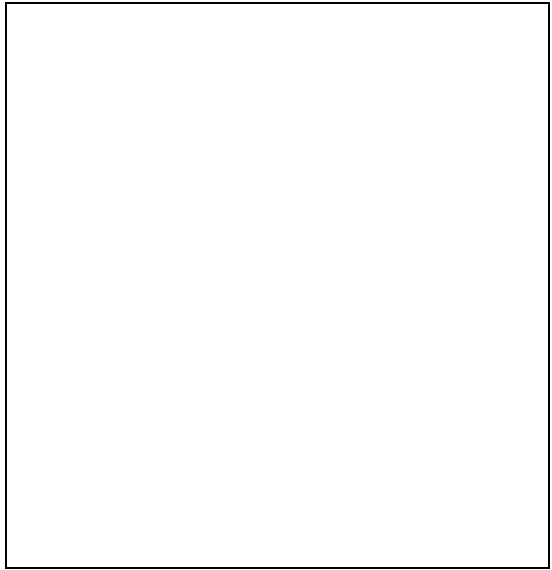

\figureheight{7.3cm}
\caption{Ground-state phase diagram of the 2D dimerized $t$$-$$J$ model 
at quarter filling obtained for the $2$$\times$$2$ unit-cell cluster.  
Values of $\tilde{J}_d$ are indicated as a contour (dashed line).  
We put some values of the spin gap $\Delta_{s}/t_2$ in the singlet-pairing 
region.}
\label{fig:4}
\end{figure}

Based on the calculated results given above, we obtain the phase diagram 
of the model on the parameter space of the two dimerizations.  The result 
is shown in Fig.~4 where the horizontal axis is the $t$-dimerization 
$\tilde{t}_d$ and vertical axis is $J_2/t_2$, the measure of $J$-dimerization.  
The value of $\tilde{J}_d$ is also plotted.  
We see that when the $t$-dimerization is predominant over the 
$J$-dimerization, the system is a Mott insulator (the effective half-filled 
band with $U_{\rm dimer}$), whereas when the $J$-dimerization is predominant 
over the $t$-dimerization, the system is in a singlet-pairing (or 
superconducting) state.\cite{add:2}  The boundary between these two phases 
approaches $J_2/t_2$$=$$2$ in the limit of strong $t$-dimerization.  
This phase diagram is thus quite consistent with the expectation discussed 
in the beginning of the paper and illustrated in Fig.~1 (b).  

It should be noted here that the phase diagram is obtained without 
finite-size scaling analysis of small-cluster data, as it is not 
feasible in any 2D-model calculations. 
We point out however that almost the same results are obtained for 
the $2$$\times$$2$ and $1$$\times$$4$ unit-cell clusters, that indications 
for the existence of the critical dimerization strength are robust, 
and that the proposed physical picture is quite obvious.  We also note that, 
in the 1D model where a finite-size scaling can be done, we have 
confirmed the same mechanism for the insulator-metal transition actually 
works.\cite{rf:15}  We therefore believe that the numerical results 
presented in this paper reflects the reality of the model at least 
qualitatively.  

Finally, let us examine the recent experimental data and some other 
calculations by taking $\kappa$-(BEDT-TTF)$_2$X as an example and 
see if we can find any relevance to our model system.  
In $\kappa$-(BEDT-TTF)$_2$Cu(NCS)$_2$, the hopping integrals $t_1$ and $t_2$ 
for holes are reported to be $t_1$$\simeq$$0.1$ eV and $t_2$$\simeq$$0.25$ eV 
from the extended H\"uckel band calculation,\cite{rf:16} which leads to the 
strength of $t$-dimerization $\tilde{t}_d$$\simeq$$1.5$.  The Coulomb 
interaction $U$ on a BEDT-TTF molecule is reported to be $\sim$$1.0$ eV, 
which is deduced from the analysis of the pressure dependence of the Knight 
shift based on the random-phase approximation.\cite{rf:17}  Thus the Coulomb 
interaction may be 4$-$10 times larger than the hopping integrals.  
Then if we assume that the exchange interactions can be estimated by the 
perturbation expression $J$$=$$4t^2/U$ we have $J_1/t_1$$\simeq$$0.4$ and 
$J_2/t_2$$\simeq$$1.0$.  It follows then that the obtained point 
$(\tilde{t}_d,J_2/t_2)$ falls on the Mott-insulating region of the phase 
diagram of Fig.~4.  Now it is interesting to note that the effect of pressure 
is to decrease $t_2/t_1$ (and simultaneously increase $J_2/J_1$);  an 
optimistic expectation would thus be that the system which is located in the 
Mott-insulating phase at ambient pressure shifts to the upper-right 
direction in Fig.~4 and can be in the singlet-pairing (or superconducting) 
phase with increasing pressure, as it actually is in experiment.  
It is also interesting to note that 
the critical dimerization strength $(J_2/t_2)_c$$\simeq$$1.7$ obtained at 
$\tilde{t}_d$$=$$1.5$ (see Fig.~4) is roughly consistent with the 
Hartree-Fock estimate\cite{rf:11} of the critical repulsion 
$(U_{\rm dimer}/t_2)_c$$\simeq$$1.0$ in the dimerized Hubbard 
model because, if we use the expression $U_{\rm dimer}$$=$$2t$$-$$J$ of an 
isolated $t$$-$$J$ dimer, we have $(J_2/t_2)_c$$\simeq$$1.0$.  
Further studies will however be required to clarify whether there is any 
continuity between the present dimerized $t$$-$$J$ model and more realistic 
dimerized Hubbard models, and also whether such large $J/t$ values are 
realized in the organic compounds $\kappa$-(BEDT-TTF)$_2$X.  

In summary, we have examined the 2D dimerized $t$$-$$J$ model at quarter 
filling by using an exact-diagonalization technique on small clusters and 
have shown that there is a critical dimerization strength which divides 
the ground-state phase of the model into the Mott-insulating phase and 
singlet-pairing phase; either of the two phases is realized depending 
on the strength of $t$- and $J$-dimerizations.  We have thereby obtained 
the ground-state phase diagram of the model in the parameter space of 
the two types of dimerization.  

This work was supported in part by Grant-in-Aid for Scientific 
Research from the Ministry of Education, Science, and Culture of 
Japan.  Financial supports of S.~N. by Sasakawa Scientific Research 
Grant from the Japan Science Society and of Y.~O. by Iketani Science 
and Technology Foundation are gratefully acknowledged.  
Computations were carried out in Computer Centers of the Institute for 
Solid State Physics, University of Tokyo and the Institute for 
Molecular Science, Okazaki National Research Organization.

\end{document}